# Forensic Scientometrics - An emerging discipline to protect the scholarly record

Leslie D. McIntosh[1], Cynthia Hudson Vitale[1,2]
[1]Digital Science; [2]Association of Research Libraries

*Corresponding author:
Leslie D. McIntosh
Email: leslie@digital-science.com

**Author Contributions:** LM and CHV wrote the article

## Abstract

Forensic Scientometrics (FoSci) is emerging as a vital discipline at the intersection of scientific integrity and security. Scholarship and scholarly communication are critical for maintaining scientific integrity, influencing public trust in science, health, technology, policy, and law. Yet, these foundations are threatened by the misuse of scientific research for personal, commercial, ideological, and geopolitical gains, including questionable practices and misconduct. The rise of paper mills and predatory publishers, along with ideological and geopolitical motivations, undermines academic integrity. This field pioneers the integration of traditional scientometric methods with ethics to address pressing challenges in research integrity and security, crucial in an era of heightened scrutiny over science's reliability. FoSci's development signifies a collective commitment to maintaining scientific trust, marked by a call for official recognition and support from stakeholders across the scientific ecosystem.

## Introduction

Moving from the 17th century to modern times, scholarly communication and research have changed. The aggregation of vast amounts of publicly available data coupled with fast and open means to analyze and interpret data and information requires the reevaluation of how research and research communication are trusted. During this time, computational power has grown and overpowered the wonders of the common and scientific worlds. Ease of transport and online communication and collaboration have offered a rise to globalization. World events also influence who moves into research and, consequently, our thinking. Science has also shifted, and instances of it being undermined have risen. Science has moved toward opening up access to research, which in turn has led ways to showcase and increase both the positive impact of good research as well as nefarious activities that threaten research.

As societal, cultural, and intellectual needs progress, so do the needs to extend and focus fields of research. A new academic discipline often emerges either in response to changes in society, technology or in research itself.  Another catalyst can be the interaction of different overlapping or non-overlapping fields.  Recognizing a new distinct area of study has emerged to respond to lack of adequate coverage by existing disciplines, communities of scholars, or societal and technological changes is often the final stage in the initial development of a field. We consider a discipline to be defined by having some of these characteristics: a specific research focus, unique specialized knowledge, organizing theories and concepts, specialized terminology, and tailored research methods (Krishnan 2009).

In this paper, we contend that while the existing fields of study of scientometrics and research ethics play crucial roles, they fall short of fully addressing the complexities surrounding the integrity and security of scientific research and communication.

To bridge this gap, we propose the concept of Forensic Scientometrics (FoSci)—a field dedicated to the critical examination and safeguarding of the scientific process and its credibility. Forensic Scientometrics stands at the intersection of analyzing scientific output and ensuring its integrity, utilizing methodologies from traditional scientometrics but with a sharper focus on the detection, mitigation and resolution of issues compromising research integrity and security. In an age where the integrity and security of science is under intense examination, the establishment and development of FoSci are imperative for reinforcing trust within the scientific community and with the public at large, employing advanced scientometric techniques to achieve these ends.

## The Evolution to Forensic Scientometrics

To grasp the need for a new field, it is important to comprehend the current landscape. Although FoSci is and will continue to be highly interdisciplinary, our argument for establishing a new discipline primarily builds upon two existing fields.

The field of **scientometrics**, while difficult to define, in this context may generally be considered the science of measuring and analyzing science itself through the use of mathematical and statistical methods (Mingers 2015).  As a field of study, scientometrics has a close disciplinary relationship with other fields, including science of science, metascience, and bibliometrics, to name a few. While the field of scientometrics emerged from key developments in the 1960s, the conceptual roots extend further back, involving the quantitative study of science, technology, and innovation processes. In its current practice, scientometrics involves the quantitative analysis of scientific publications and research outputs in this larger context. It encompasses the measurement and evaluation related to scientific activities, such as the impact of research, patterns of collaboration among researchers, citation analysis, and the productivity assessment and influence of individuals, institutions, or scientific journals. Scientometrics employs statistical and mathematical methods to derive meaningful insights into the structure and dynamics of scientific knowledge, contributing to our understanding of the scientific community's development and impact over time. Yet, it primarily focuses within the context of the scientific ecosystem and does not include how science is communicated or interpreted.

**Research ethics** also plays a role primarily within science rather than being translated outside of academic and research institutions. Dimensions of research ethics have been suggested to include (DuBois 2018): normative ethics, compliance with regulations, rigor and reproducibility of science, social value, and workplace relationships. Research ethics is a vital field that ensures the integrity and validity of scientific inquiry by upholding ethical principles such as respect for persons, beneficence, and justice in the conduct of research. This discipline safeguards the rights and welfare of human and animal subjects, mandates informed consent, and promotes the fair distribution of research benefits and burdens. Guided by ethical guidelines and regulated by institutional review boards (IRBs), research ethics emphasizes the responsible use of resources, the management of conflicts of interest, and professionalism in research for ensuring good research and addressing the concerns of stakeholders.

## Key Developments and Initiatives

Scientometrics, research ethics, and research integrity have evolved and expanded over the years. Additionally, the emergence of Forensic Scientometrics (FoSci) has been influenced by other factors, including computational advancements and globalization. Although not exhaustively listed, these elements are crucial to consider in understanding the development of the FoSci field. Below is an overview of key events.

- **1955**: Creation of the **Impact Factor (Garfield 2006)** to help select journals for the Science Citation Index (SCI).
- **1963**: Publication of Derek J. de Solla Price's book **"Little Science, Big Science"**, marking a foundational moment in the study of scientometrics (Price 1963).
- **Late 1960s**: Scientometrics is increasingly recognized as a distinct field of study, with the term likely being used in articles and discussions among scholars.
- **1978**: Launch of the **Journal of Scientometrics**, significantly contributing to the formalization and development of the field.
- **1978**: The **International Committee of Medical Journal Editors (ICMJE)**, though formed earlier, becomes increasingly influential in promoting ethical standards in medical research publishing.
- **1980s-1990s**: Research integrity begins forming as a recognized area, parallel to developments in scientometrics, focusing on the ethical aspects of conducting and reporting research.
- **1997**: Formation of the **Committee on Publication Ethics (COPE)**, providing guidelines and standards for addressing ethical issues in publication ((COPE) No year).
- **2005**: Establishment of the **World Conference on Research Integrity (WCRI)**, further solidifying the global commitment to research integrity.
- **2010**: Launch of **Retraction Watch**, a blog tracking retractions of scientific papers, highlighting issues of research integrity.
- **Early 2010s Onwards**: Increasing **awareness on social media** about research practices and integrity.
- **2018**: The **Retraction Watch Database** is launched, providing a searchable database of retracted papers, further enhancing transparency in the scientific community.

- **2018 Onwards**:
    - 'Sleuthing' and 'Sleuths' gains consistency in describing individuals who find problems within scholarly publications (McCook 2018)
    - Research Integrity and manipulation of science become part of mainstream media conversations
    - Research Security emerges in government policies and in scientific literature (Lewandowsky 2023)

## Nefarious Activities and the Need for a New Discipline

Trusted scholarship and scholarly communication are pivotal in preserving the integrity of science, essential for supporting confidence in scientific discoveries, health choices, technological advancements, policy creation, and legal discourse. However, this foundation is compromised by the exploitation of scientific endeavors for diverse gains.

The **motivations** for engaging in unethical research practices are multifaceted, spanning individual, ideological and geopolitical strategies. At the **individual level**, perverse incentives such as career advancement can drive researchers towards questionable research practices (QRP) and misconduct. **Commercial interests and ideology** drive an intentional or opportunistic usurping of the norms in conducting and communicating research. **Geopolitical interests**, those of state-led actors, also drive research and technology activities as seen in the cited report (Braw 2022).

The rapid spread of **means** to share what looks like legitimate scientific research is alarming. Paper mills - enterprises that sell authorship without requiring actual research contribution (COPE 2022)- represent a significant threat. Similarly, predatory publishers and conferences exploit researchers by charging fees to present or publish work without proper vetting (Linacre 2022). The creation of journals aimed at promoting specific industry or ideological outcomes under the guise of peer-reviewed research also works as an enterprise to undermine the integrity of academic inquiry. Additionally, the recruitment of researchers to further ideological, commercial or geopolitical interests under scientific pretenses is concerning (Drope 2001, Michaels 2020).

**Impact of Malpractices on the Scientific Ecosystem** These personal motivations can have career and organizational consequences. Despite arguments that legitimate researchers may not engage with such fraudulent outputs, their mere existence undermines the integrity of the scientific record and facilitates the spread of misinformation. As people are promoted under false credentials (e.g., publications without the requisite knowledge) they may move into decision-making positions but not truly be experts. Additionally, university rankings will be falsely influenced by substanceless metrics (Ansede 2023,ROARS 2024). Individuals who compromise their research through QRP and misconduct also become vulnerable to outside influence (e.g., targeting to push political agendas) thus blurring the lines between scholarly pursuit and political activism or geopolitical influence.

Such a landscape described in this section necessitates the emergence of FoSci as an essential discipline aimed at preserving the sanctity of academic discourse. Through the vigilant examination and safeguarding of the scholarly record, FoSci seeks to mitigate internal and external threats to scientific integrity and security, ensuring the continuity of genuine and trusted scientific advancement.

## Introducing Forensic Scientometrics (FoSci)

Forensic Scientometrics (FoSci) emerges as a pivotal discipline aimed at uncovering, analyzing and addressing unethical practices within scholarly communication, including research misconduct, data fabrication, plagiarism, as well as other forms of academic dishonesty. From the onset, it also involves and incorporates broader contexts of how science is used and abused from actors within and outside of traditional research institutions.

Its core mission extends to identifying patterns, trends, and predictors of misconduct both within and external to the scientific community, through an analysis of scientific literature and science communication channels. FoSci is dedicated to assessing the impact of such unethical behaviors on the scientific community, the integrity of the research record, and the public's trust in science. This involves a diverse array of detection and analysis techniques, including the examination of authorship networks, retracted articles, nefarious articles and publication practices, citation networks, language manipulation, and the analysis of social media to understand the misuse of scientific information and to scrutinize social media patterns themselves.

### What's in a Name

The process of naming a discipline encapsulates the ambition, direction, and uniqueness of the field, embodying its potential to contribute to knowledge, solve societal problems, and open new avenues of research and development. In essence, a name not only defines the field but also empowers it, enabling stakeholders to rally around a common cause and identity, thereby facilitating its growth and the achievement of its goals.

The name has been chosen due to the investigative nature coupled with the methodologically rich approach of the discipline.

Forensics refers to applying scientific knowledge and methods to matters of law, particularly for investigating crimes and providing evidence in legal proceedings. Yet, colloquially, it means something formal that is investigative in nature. While traditionally 'forensics' encompasses fraud investigations and syndicate networks encompass legal realms, the intentional manipulation of anything scientific (i.e., the process of scientific discovery or manipulation) does not have a special field or discipline.

Scientometrics, as already discussed, brings in a methodological angle to the discipline and is being used to encapsulate the principles of bibliometrics, the science of science, and metascience. Moreover, 'forensic science' already exists as a distinct and different discipline from FoSci.

## Theory and Practice

Applied fields frequently adopt an interdisciplinary approach, integrating insights from multiple scientific disciplines to tackle real-world problems. The knowledge transfer – from identifying practical issues (e.g., paper mills) to devise solutions and explore their theoretical foundations will play a crucial role in the field of forensic science (FoSci).

Both academics and practitioners are engaged in scrutinizing anomalies in research practices and communication. These endeavors, intrinsic to both the academic sphere and professional domains, mirror the dynamics observed in other practice-based disciplines, such as pathology, where notable distinctions exist between academic research and field application. As depicted in Table 1, practitioners primarily focus on the principles and methodologies for detecting the origins and methods of manipulated scientific data, potentially providing expert testimony. In contrast, academics are more inclined toward conducting research, advancing education, and enhancing the understanding of research integrity and security.

Despite these differences, natural intersections emerge, with practitioners often engaging in research activities and academics participating in applied practices.

**Table 1: Focus, objectives, work environment and engagement for the proposed Forensic Scientometric discipline.**

|  | Theory | Practice |
|---|---|---|
| Focus | Research, education, and knowledge dissemination | Application of scientometric principles and techniques, research management, and information management |
| Objectives | Advance scientific understanding, develop new methodologies, and train future researcher | Determine the cause and manner manipulation; provide expert testimony in ethical and legal proceedings |
| Work Environment | <ul><li>Universities</li><li>Research institutions</li><li>Government</li></ul> | <ul><li>Institutional administration Publishers</li><li>Government and funders</li><li>Commercial entities engaged with research</li><li>Independent consultants</li></ul> |
| Engagement | Engage in conducting research studies, publishing academic papers, teaching students, and participating in academic conferences and professional associations. | Engage in investigations into nefarious or suspicious activities. They participate in professional association conferences. |

## Tiers of Specializations

A hallmark of applied emerging fields are specializations, which have been observed in forensic scientometrics. Based on current trends, the multi-faceted practice of FoSci has three tiers of specialization: macro, meso, and micro. The macro tier pertains to investigating the process flow of integrity and security in and out of the scientific realm. For example, this will include scientific communication in the media and science applications in law, political uses, or abuses of science. At the meso tier lies investigation into science communication operations and authorship networks with the heaviest reliance on bibliometric databases that connect publications with grants with authors and other artifacts of research. The micro tier practitioners delve into granular checks of the research presented such as the data, text analysis, chemicals, and image manipulation within publications. Table 2 provides sample specializations with examples.

**Table 2: Suggested specializations in existence within Forensic Scientometrics**

|       | Specializations | Examples |
|-------|-----------------|----------|
| Macro | Journalistic coverage<br>Publishing standards | Retraction Watch (Oransky 2023) |
| Meso  | Authorship Networks<br>Citation Cartels | Authorship-for-Sale (Porter 2024) |
| Micro | Image manipulation<br>Text manipulation<br>AI detection | Image Duplication (Bik 2016)<br>Problematic Paper Screener (Cabanac 2022) |

## Forensic Scientometrics as a Discipline

As a discipline (Tight 2020), we can expect to explore the theoretical underpinnings of a field in addition to the applied learnings. We suggest the following components of a discipline wrapped with two supporting, vital parts of ethics and professional development:

**Investigations -** Conducting case studies, observational research, and formal investigations to examine patterns of trust and mistrust in scientific communication practices, identify anomalies or irregularities that may suggest misconduct, and assess the impact of such behaviors on scientific integrity, security and public trust in science.

**Pedagogy** - Developing educational programs, workshops, and training materials to raise awareness about research integrity and security, ethical practices in scientific communication, and the principles and applications of Forensic Scientometrics among researchers, students, and professionals.

**Methodology Development** - Designing and refining statistical and computational methods, algorithms, and tools for scrutinizing scientific publications, citation networks, authorship patterns, and other bibliometric indicators to detect potential research misconduct, data fabrication, plagiarism, or other unethical practices.

**Research** - developing and refining models and methods of the field, developing policy. Research involves the systematic investigation of phenomena, theories, or problems within the field. Research activities may include conducting experiments, collecting and analyzing data, developing theories, and publishing findings in academic journals.

**Impact Assessment** - Evaluating the consequences of misconduct, retractions and disinformation on the scientific community and public. This will include a wide range of topics such as studying the impact on citation patterns, analyzing public trust in science, and the effects of erroneous finding propagation.

**Theory Development** - Developing theoretical frameworks and models to understand the underlying factors, motivations, and mechanisms that contribute to various forms of research misconduct, as well as the broader implications for the scientific ecosystem.

**Interdisciplinarity** - Fostering interdisciplinary collaborations with other fields, such as ethics, sociology, psychology, information science, business, law, intelligence, and computer science, to integrate diverse perspectives and expertise in addressing complex issues related to research integrity and scientific communication.

**Policy Development** - Contributing to the formulation of policies, guidelines, and best practices for promoting research integrity, preventing misconduct, and addressing ethical violations in scientific publishing and communication, in collaboration with relevant stakeholders and organizations.

—

**Ethics and Standards** - Establishing ethical standards, codes of conduct, and best practices for the responsible and ethical conduct of Forensic Scientometrics activities, in alignment with existing guidelines from organizations like the Committee on Publication Ethics (COPE) and the International Committee of Medical Journal Editors (ICMJE).

**Professional Development**: These organizations may offer conferences, workshops, publications, and networking opportunities to help individuals stay current with developments in the field and enhance their skills and knowledge.

## Current Challenges and Opportunities

Before advancing FoSci as an academic discipline several challenges and opportunities must be considered to advance and improve science:

### Cleaning up Science

The emerging field of FoSci has much inquiry and work to catch up on. The growth of paper mills and citation networks coupled with the ease of publishing text online has facilitated the explosion of questionable and fraudulent science. The journey ahead for all working in or near scholarship and scholarly communications involves confronting some uncomfortable realities about the scientific landscape. Acknowledging the need for a systematic cleanup of the scientific corpus is paramount. This endeavor is not just about identifying and rectifying instances of subpar science or outright fraudulent activities, though these are significant challenges. It also involves recognizing that not all institutions have the resources or the willingness to address these issues. As a community, we must grapple with these disparities and work towards equitable solutions that uphold the integrity of science across all research ecosystems.

Norms and practices for the reporting of lapses in research integrity and security in an ethical and responsible manner have not yet been well established across the FoSci community. While COPE and other ethical organizations have developed practices for publishers and journal editors to follow when a letter of concern is submitted, guidelines for researchers who discover issues in the scientific literature are limited. While the mechanism for submitting a letter of concern is straightforward, the more ethical considerations around what types or levels of research integrity lapses to report and their appropriate remediation have not been well established.

## Growing and Training the Community

There are few appropriate formal and informal communication channels to report, discuss, and interpret probable research integrity lapses. Through the formalization of a field, modes of communication will continue to develop and grow, such as the establishment of dedicated conferences and the founding of specialized journals that will allow the broad sharing of practices and methods.

As a research method that is critical for scientometrics, bibliometric analysis among researchers is not a well-established competency or skill. Bibliometric analysis, which involves the quantitative study of scientific publications and their impact, requires a deep understanding of statistical methods, data visualization techniques, and the ability to interpret complex datasets. Additionally, expertise in handling bibliographic databases and familiarity with various bibliometric indicators and software tools is essential. Unfortunately, many individuals in the academic and research communities lack this specialized knowledge and training. This skill gap hinders the effective application of scientometric methods to assess research trends, evaluate scientific output, and inform policy-making in the research environment.

Perhaps even more challenging is that with the advent of new, more accessible artificial intelligence and machine-learning technologies, new modes of manipulating science have been used that are almost imperceptible to all but the researcher or author. As a field, the FoSci community would benefit from collective advocacy for the increased transparency of the use of AI and other technologies as well as improved capabilities within the community.

Looking ahead, we anticipate both a need for new specializations. For example, that between the intersection of science and law. Scientific publications are increasingly being introduced in legal proceedings as evidence. However, the critical examination of these papers—their validity, reliability, and relevance outside the academic sphere—often goes unaddressed. This gap highlights a significant area where expert forensic scientometricians can contribute, ensuring that the science underpinning legal arguments is both robust and relevant.

## Improving Incentives and Celebrating Successes

While research ethics and integrity are fundamental to the veracity of the scientific process, they do not receive the same recognition or reward as other research achievements. The current academic reward system focuses on publications, grant awards, and citations, which can lead researchers to prioritize these metrics over ethical considerations. As a result, there is a need for a shift in the academic culture and reward structure to value and incentivize ethical research practices. This could include recognizing this work in promotion and tenure decisions, providing funding for research that leverages forensic scientometric practices, and incorporating bibliometric analysis and scientometric training into research education programs. By creating a more balanced environment, we can encourage researchers to prioritize ethical research practices, ultimately leading to more trustworthy and impactful research outcomes.

Equally important is the celebration of science's successes and its inherent quest for knowledge, even when that quest leads to questions rather than answers. The ability to question, to doubt, and to be wrong is a fundamental part of the scientific process. It is through this rigorous examination and re-examination of our hypotheses that we inch closer to the truth. FoSci can play a critical role in highlighting and supporting these aspects of scientific inquiry, reinforcing the value of transparency and integrity in all scientific endeavors.

# Discussion

The exploration of forensic scientometrics reveals a critical landscape where the stakes of scientific integrity and security intersect with individual, ideological, and geopolitical motivations. As we navigate the evolving landscape of scientific inquiry, the emergence of forensic scientometrics as a distinct field reflects a collective commitment to upholding research integrity and security.

We argue that aiming an emerging discipline serves as a foundational element for developing methods, solving problems, garnering support, and building a cohesive community. The act of naming is integral to legitimizing and acquiring resources for the discipline, marking an essential step in their evolution and impact on society.

And despite the importance of scientometrics and research ethics, they do not entirely cover the complexities, communication, and information propagation concerning the integrity and security of scientific research. The suggested field of FoSci merges traditional scientometric methods with ethics and a strong emphasis on identifying, addressing, and solving challenges to research integrity and security, which is vital in today's climate of scrutiny over science's reliability. By enhancing scientific trust and employing sophisticated scientometric tools, FoSci is essential for the scientific community and public confidence.

We explored the evolution of scientometrics, ethics, and research integrity, demonstrating their limitations for our objectives. Our analysis extended to the misuse of scientific research for various benefits, highlighting the vulnerabilities of scholarly communication. We introduced Forensic Scientometrics (FoSci) as a practical, interdisciplinary field that combines insights from several domains to address real-world issues, such as academic fraud and unethical practices like paper mills, authorship-for-sale, image manipulation, and plagiarism. This specialization, as seen in practitioners who investigate these issues, underscores the emergence of FoSci as a distinct academic discipline, characterized by its specific focus, methodologies, and the crucial role of knowledge transfer in addressing and understanding scientific malpractices.

From the pioneers who have tirelessly exposed misconduct to the institutional changes taking place, the journey towards a recognized field is well underway. It is now time to define the field to support and grow the community of professionals working to uphold the scaffolds of scientific trust.

In conclusion, Forensic Scientometrics (FoSci) stands at the precipice of becoming a transformative force within and for science, poised to ensure the integrity and trustworthiness of scholarly endeavors. Our discourse not only illuminates the existence of a burgeoning community dedicated to safeguarding the bedrock of scientific research but also underscores the paramount importance of this discipline in defending the scholarly record from the myriad threats it faces. As we stand on the cusp of officially recognizing Forensic Scientometrics, we issue a call to all stakeholders—researchers, funding bodies, institutions, and policymakers—to join forces in nurturing, funding, and championing this critical field. It is through our collective endeavor that FoSci can flourish, transforming from a budding area of study into a cornerstone of scientific integrity.

# Acknowledgements

Sincere thanks to Hélène Draux for assistance refining bibliometric queries, Simon Linacre, and Daniel Hook for edits and discussions.

# Funding Statement

No external funding was provided for this work.

# Data Availability Statement

No data were used for this paper.

## Conflicts of Interest

Leslie McIntosh is a full-time employee and Vice President of Research Integrity at Digital Science. Cynthia Hudson Vitale receives a part-time salary from Digital Science. Both provide advice on products related to research integrity.

## References

Ansede, M. Spanish university administrator and colleagues linked to 'factory' producing fraudulent scientific studies. *El País* (2023).

Bik, E. M., Casadevall, A. & Fang, F. C. The Prevalence of Inappropriate Image Duplication in Biomedical Research Publications. *mBio* **7**, e00809-16 (2016).

Braw, E. *In-Depth Briefing #33: Research Security: A New Frontier*. (2022).

Cabanac, G., Labbé, C. & Magazinov, A. The "Problematic Paper Screener" automatically selects suspect publications for post-publication (re)assessment. *arXiv* (2022) doi:10.48550/arxiv.2210.04895.

COPE. COPE History Timeline. https://publicationethics.org/about/history.

COPE & STM. Paper Mills — Research report from COPE & STM. (2022) doi:10.24318/jtbg8ihl.

Drope, J. & Chapman, S. Tobacco industry efforts at discrediting scientific knowledge of environmental tobacco smoke: a review of internal industry documents. *J. Epidemiology Community Heal.* **55**, 588 (2001).

DuBois, J. M. & Antes, A. L. Five Dimensions of Research Ethics. *Acad. Med.* **93**, 550–555 (2018).

Garfield, E. The History and Meaning of the Journal Impact Factor. *JAMA* **295**, 90–93 (2006).

Krishnan, A. What are academic disciplines? Some observations on the disciplinarity vs. interdisciplinarity debate. (2009).

Lewandowsky, S. *et al.* Misinformation and the epistemic integrity of democracy. *Curr. Opin. Psychol.* **54**, 101711 (2023).

Linacre, S. *The Predator Effect*. (Against the Grain, 2022). doi:10.3998/mpub.12739277.

McCook, A. Philosophers, meet the plagiarism police. His name is Michael Dougherty. https://retractionwatch.com/2018/06/12/philosophers-meet-the-plagiarism-police-michael-dougherty/ (2018).

Michaels, D. *Triumph of Doubt*. (Oxford University Press, 2020).

Mingers, J. & Leydesdorff, L. A review of theory and practice in scientometrics. *Eur. J. Oper. Res.* **246**, 1–19 (2015).

Oransky, I. & Marcus, A. There's far more scientific fraud than anyone wants to admit. (2023).

Porter, S. J. & McIntosh, L. D. Identifying Fabricated Networks within Authorship-for-Sale Enterprises. *arXiv* (2024) doi:10.48550/arxiv.2401.04022.

Price, D. J. D. S. *Little Science, Big Science*. (Columbia University Press, 1963). doi:10.7312/pric91844.

ROARS. The Top Italian Scientists and their Journal. https://www.roars.it/the-top-italian-scientists-and-their-journal/ (2024).

Tight, M. Higher education: discipline or field of study? *Tert. Educ. Manag.* **26**, 415–428 (2020).